\documentclass[12pt]{article}
\usepackage[english,german,french,polish]{babel}
\usepackage[T1]{fontenc}

\begin{document}

\baselineskip 0.72cm
\topmargin -0.4in
\oddsidemargin -0.1in

\let\ni=\noindent

\renewcommand{\thefootnote}{\fnsymbol{footnote}}

\newcommand{\Stk}{SuperKamiokande }

\pagestyle {plain}

\setcounter{page}{1}

\pagestyle{empty}



~~~~~
\begin{flushright}
IFT-02/02
\end{flushright}

\vspace{0.2cm}

{\large\centerline{\bf Neutrino bimaximal mixing}}
{\large\centerline{\bf and unitary deformation of fermion universality{\footnote {Supported in part by the Polish State Committee for Scientific Research (KBN), grant 5 P03B 119 20 (2001--2002).}}}}

\vspace{0.5cm}

{\centerline {\sc Wojciech Kr\'{o}likowski}}

\vspace{0.23cm}

{\centerline {\it Institute of Theoretical Physics, Warsaw University}}

{\centerline {\it Ho\.{z}a 69,~~PL--00--681 Warszawa, ~Poland}}

\vspace{0.5cm}

{\centerline{\bf Abstract}}

\vspace{0.3cm}

An effective texture is presented for six Majorana neutrinos, three active and three (conventional) sterile, based on a $6\times 6$ mass matrix whose $3\times 3$ Dirac component ({\it i.e.}, active--sterile component) is conjectured to get a hierarchical {\it fermion universal form}, similar to the previously constructed $3\times 3$ mass matrices for charged leptons as well as  for up and down quarks. However, for neutrinos this form becomes {\it unitarily deformed} by the action of {\it bimaximal mixing}, specific in their case. The $3\times 3$ lefthanded and righthanded  components ({\it i.e.}, active-active and sterile-sterile components) of the $6\times 6$ mass matrix are diagonal with degenerate entries of opposite sign. They dominate over the $3\times 3$ Dirac component. In such a texture the neutrino masses are  $ m_1 = -m_4 \simeq m_2 = -m_5  \simeq m_3 = -m_6 $ with the mass-squared differences $\Delta m^2_{21} \ll \Delta m^2_{32} \simeq \Delta m^2_{31}$ and $\Delta m^2_{41} = \Delta m^2_{52} = \Delta m^2_{63} = 0$. The last equality implies in our texture the absence of oscillations for three (conventional) sterile neutrinos. Thus, these neutrinos are here effectively decoupled, what is realized evidently in another way than through the popular seesaw mechanism. There remain the oscillations of three active neutrinos, getting the form as for bimaximal mixing, but with the mass spectrum following from our texture.

\vspace{0.3cm}

\ni PACS numbers: 12.15.Ff , 14.60.Pq , 12.15.Hh .

\vspace{0.6cm}

\ni January 2002

\vfill\eject

~~~~~
\pagestyle {plain}

\setcounter{page}{1}

\ni {\bf 1. Introduction. } As is well known, three Dirac neutrinos are $ \nu^{(D)}_\alpha = \nu_{\alpha L} + \nu_{\alpha R}\;\;(\alpha = e\,,\, \mu\,,\, \tau)$, while three Majorana active neutrinos and three Majorana (conventional) sterile neutrinos become $\nu^{(a)}_\alpha = \nu_{\alpha L} + \left( \nu_{\alpha L} \right)^c $ and $\nu^{(s)}_\alpha = \nu_{\alpha R} + \left( \nu_{\alpha R} \right)^c\;\;(\alpha = e\,,\, \mu\,,\, \tau)$, respectively. The neutrino mass term in the Lagrangian gets generically the form

\vspace{-0.2cm}

\begin{equation} 
- {\cal L}_{\rm mass} = \frac{1}{2}\sum_{\alpha \beta} (\overline{\nu_\alpha^{(a)}} \,,\, \overline{\nu_{\alpha}^{(s)}}) \left( \begin{array}{cc} M^{(L)}_{\alpha \beta} & M^{(D)}_{\alpha \beta} \\ M^{(D)*}_{\beta \alpha}  & M^{(R)}_{\alpha \beta} \end{array} \right) \left( \begin{array}{c} \nu^{(a)}_\beta \\ \nu^{(s)}_{\beta} \end{array} \right) \;.
\end{equation} 
 
\ni If $M^{(L)}_{\alpha \beta} $ and $M^{(R)}_{\alpha \beta} $ are not all zero, then in nature there are realized six Majorana neutrino mass fields $\nu_i $ or states $|\nu_i \rangle \;\;(i = 1,2,3,4,5,6)$ connected with six Majorana neutrino flavor fields $\nu_\alpha $ or states $| \nu_\alpha\rangle \;\;(\alpha = e\,,\, \mu\,,\, \tau\,,\,e_s\,,\, \mu_s\,,\, \tau_s)$ through the unitary transformation

\vspace{-0.2cm}
 
\begin{equation} 
\nu_\alpha  = \sum_i U_{\alpha i}  \nu_i \;\;{\rm or}\;\; |\nu_\alpha \rangle = \sum_i U^*_{\alpha i} |\nu_i \rangle \;,
\end{equation} 

\ni where we passed to the notation $\nu_\alpha \equiv \nu^{(a)}_\alpha $ and $\nu_{\alpha_s} \equiv \nu^{(s)}_\alpha $ for $\alpha = e\,,\, \mu\,,\, \tau$. Of course, $ \nu^{(a)}_{\alpha L} = \nu_{\alpha L} $, $\nu^{(a)}_{\alpha R} = (\nu_{\alpha L})^c $ and $\nu^{(s)}_{{\alpha} R} = \nu_{\alpha R}$, $\nu^{(s)}_{\alpha L} = (\nu_{\alpha R})^c $ for $\alpha = e\,,\, \mu\,,\, \tau$. Thus, the neutrino $6\times 6$ mass matrix $M = (M_{\alpha \beta})\;\;(\alpha, \beta = e\,,\, \mu\,,\, \tau\,,\,e_s\,,\, \mu_s\,,\, \tau_s)$ is of the form 

\vspace{-0.2cm}

\begin{equation}
M = \left( \begin{array}{cc} M^{(L)} & M^{(D)} \\ M^{(D)\dagger}  & M^{(R)} \end{array} \right) \;.
\end{equation}

\ni The neutrino $6\times 6$ mixing matrix $U = (U_{\alpha i})\;\;(i = 1,2,3,4,5,6)$ appearing in Eqs. (2) is, at the same time, the unitary $6\times 6$ diagonalizing matrix,

\vspace{-0.2cm}

\begin{equation} 
U^\dagger M U = M_{\rm d} \equiv {\rm diag}(m_1\,,\,m_2\,,\,m_3\,,\,m_4\,,\,m_5\,,\,m_6)\;,
\end{equation} 

\ni if the representation is used, where the charged-lepton $3\times 3$ mass matrix is diagonal. This will be assumed henceforth.

\vspace{0.2cm}

\ni {\bf 2. Model of neutrino texture. } In the present paper we study the model of neutrino texture, where the $3\times 3$ submatrices in Eq. (3) are

\vspace{-0.2cm}
 
\begin{eqnarray} 
M^{(L)} & = & {\stackrel{0}{m}} \left( \begin{array}{ccc} 1 & 0 & 0 \\ 0 & 1 & 0 \\ 0 & 0 & 1 \end{array} \right) = - M^{(R)} \,, \nonumber \\ M^{(D)} & = &  {\stackrel{0}{m}} \left( \begin{array}{ccc} \frac{1}{\sqrt{2}} & \frac{1}{\sqrt{2}} & 0 \\ -\frac{1}{2} &  \frac{1}{2} & \frac{1}{\sqrt{2}} \\ \frac{1}{2} & -\frac{1}{2} & \frac{1}{\sqrt{2}}  \end{array} \right) \, \left( \begin{array}{ccc} \tan 2\theta_{14} & 0 & 0 \\ 0 & \tan 2\theta_{25} & 0 \\ 0 & 0 & \tan 2\theta_{36} \end{array} \right)
\end{eqnarray}

\ni with ${\stackrel{0}{m}}> 0 $ being a mass scale and $ \tan 2\theta_{i j}\;\;(i j = 14\,,\,25\,,\,36)$ denoting three dimensionless parameters. Notice that in the Dirac component $M^{(D)}$ of the neutrino mass matrix $M$ there appears the familiar matrix of bimaximal mixing that {\it deforms unitarily} a diagonal hierarchical structure ({\it cf.} Eq. (10) later on).

One can show that the unitary diagonalizing matrix $U$ for the mass matrix $M$ defined in Eqs. (3) and (5) gets the form

\begin{equation}
U = {\stackrel{1}{U}}{\stackrel{0}{U}}\;,\;{\stackrel{1}{U}} =  \left( \begin{array}{cc} U^{(3)} & 0^{(3)} \\ 0^{(3)} & 1^{(3)} \end{array} \right) \;,\; {\stackrel{0}{U}} = \left( \begin{array}{cc} C^{(3)} & -S^{(3)} \\ S^{(3)} & C^{(3)} \end{array} \right)\;,  
\end{equation} 

\ni where

\begin{eqnarray} 
U^{(3)} =   \left( \begin{array}{ccc} \frac{1}{\sqrt{2}} & \frac{1}{\sqrt{2}} & 0 \\ -\frac{1}{2} &  \frac{1}{2} & \frac{1}{\sqrt{2}} \\ \frac{1}{2} & -\frac{1}{2} & \frac{1}{\sqrt{2}}  \end{array} \right) & , &
1^{(3)} =  \left( \begin{array}{ccc} 1 & 0 & 0 \\ 0 & 1 & 0 \\ 0 & 0 & 1 \end{array} \right) ,\nonumber \\ C^{(3)} =  \left( \begin{array}{ccc} c_{14} & 0 & 0 \\ 0 & c_{25} & 0 \\ 0 & 0 & c_{36} \end{array} \right)\;\; & , &  S^{(3)} =  \left( \begin{array}{ccc} s_{14} & 0 & 0 \\ 0 & s_{25} & 0 \\ 0 & 0 & s_{36} \end{array} \right) 
\end{eqnarray} 

\ni with$ s_{i j} = \sin \theta_{i j}$ and $ c_{i j} = \cos \theta_{i j}$, while the neutrino mass spectrum is

\begin{equation}
m_{i,j} = \pm {\stackrel{0}{m}}\sqrt{1 + \tan^2 2\theta_{i j}} 
\end{equation}

\ni and, equivalently, can be described by the equalities

\begin{equation}
\left(c^2_{i j} - s^2_{i j}\right) m_{i,j} = \pm {\stackrel{0}{m}}
\end{equation}

\ni $(i j = 14\,,\,25\,,\,36)$. The easiest way to prove this theorem is to start with the diagonalizing matrix $U$ given in Eqs. (6) and (7), and then to construct the mass matrix $M$ defined in Eqs. (3) and (5) by making use of the formula $M_{\alpha \beta} = \sum_i U_{\alpha i} m_i U^*_{\beta i}\,$, where the mass spectrum (8) or (9) is to be taken into account.

From Eqs. (6) and (7) we can see that in our texture the mixing angles give $ c_{12} = 1/\sqrt{2} = s_{12}$, $ c_{23} = 1/\sqrt{2} = s_{23} $ and $ c_{13} = 1$, $s_{13} = 0 $, whilst $ c_{ij} \,,\, s_{ij}\;\;(ij = 14\,,\,25\,,\,36)$ are to be determined from the experiment. Evidently, these values, in particular $ s_{13} = 0$, may be a subject of correction.

Note that the Dirac $3 \times 3$ component  of the neutrino mass matrix $M$ {\it transformed unitarily} by means of the factor ${\stackrel{1}{U}}$ of the diaginalizing matrix $U$ becomes diagonal and may get a {\it hierarchical} structure, 

\begin{equation} 
\left( {\stackrel{1}{U}\!^\dagger} M {\stackrel{1}{U}} \right)^{(D)} = U^{(3) \dagger}M^{(D)} = {\stackrel{0}{m}}\left(\begin{array}{ccc} \tan 2\theta_{14} & 0 & 0 \\ 0 & \tan 2\theta_{25} & 0 \\ 0 & 0 & \tan 2\theta_{36} \end{array}  \right) \;, 
\end{equation} 

\vspace{0.2cm}

\ni {\it similar} to the Dirac-type mass matrices for charged leptons as well as for up and down quarks, all dominated by their diagonal parts. The transforming factor ${\stackrel{1}{U}}$ given in Eq. (6) works effectively thanks to its $3\times 3$ submatrix $U^{(3)}$ that is just the familiar {\it bimaximal mixing matrix} [1], specific for neutrinos, describing satisfactorily the observed oscillations of solar $\nu_e$'s and atmospheric $\nu_\mu $'s. 

Specifically, the Dirac $3 \times 3$ component of the neutrino mass matrix, when the bimaximal mixing characteristic for neutrinos is transformed out unitarily, may be conjectured in a {\it fermion universal form} that was shown to work very well for the mass matrix of charged leptons [2] and neatly for  mass matrices of up and down quarks [3] (obviously, in those three cases of charged fundamental fermions there exist only Dirac-type mass matrices). Then, for neutrinos we get [4]

\begin{equation}
\left( {\stackrel{1}{U}\!^\dagger} M {\stackrel{1}{U}} \right)^{(D)}  = \frac{1}{29} \left(\begin{array}{ccc} \mu \varepsilon & 2\alpha  & 0 \\ 
2\alpha  & 4\,\mu\, (80 + \varepsilon)/9 & 8\sqrt{3}\, \alpha \\ 
0 & 8\sqrt{3}\, \alpha & 24\, \mu\, (624 + \varepsilon)/25 \end{array} \right) \;,
\end{equation}

\vspace{0.2cm}

\ni where $\mu > 0$ , $\alpha > 0$ and $\varepsilon > 0$ are some neutrino parameters. Since already for charged leptons $\varepsilon^{(e)} = 0.172329$ is small [2], we will put for neutrinos $\varepsilon \rightarrow 0$. We will also conjecture that for neutrinos $\alpha/\mu $ is negligible, as for charged leptons the small $\left( \alpha^{(e)}/\mu^{(e)} \right)^2 = 0.023^{+0.029}_{-0.025} $ [2] gives the prediction $ m_\tau = m^{\rm exp}_\tau = 1777.03^{+0.30}_{-0.26}$ MeV [5], when $m_e = m^{\rm exp}_e$ and $m_\mu = m^{\rm exp}_\mu $ are used as inputs, while with $\left( \alpha^{(e)}/\mu^{(e)} \right)^2 = 0 $ the prediction becomes $ m_\tau = 1776.80$ MeV. In such a case, from Eqs. (10) and (11) we can conclude that

\begin{equation}
{\stackrel{0}{m}}\tan 2\theta_{14} = \frac{\mu}{29} \varepsilon \rightarrow 0 \,,\,{\stackrel{0}{m}} \tan 2\theta_{25} = \frac{\mu}{29} \frac{4\cdot 80}{9}  = 1.23 \mu \,,\,{\stackrel{0}{m}} \tan 2\theta_{36} = \frac{\mu}{29} \frac{24 \cdot 624}{25} = 20.7 \mu\,
\end{equation}

\ni in Eqs.  (5), (8) and (10). Hence, from Eqs. (8)

\begin{equation} 
m_{1,4} = \pm {\stackrel{0}{m}}\,,\, m_{2,5} = \pm \sqrt{{\stackrel{0}{m}}\,\!^2 + 1.50 \mu^2} \,,\, m_{3,6} = \pm \sqrt{ {\stackrel{0}{m}}\,\!^2 + 427 \mu^2}\;.
\end{equation} 

\vspace{0.6cm}

\ni {\bf 3. Neutrino oscillations.} Accepting the formulae (12) and making tentatively the conjecture that $\mu\! \ll \!{\stackrel{0}{m}} $, we can operate with the approximation, where $0 \leq \tan 2\theta_{ij} \ll 1$ $ (ij = 14\,,\,25\,,\,36)$. Then, we get the case of nearly degenerate spectrum  $m_1 \simeq m_2 \simeq m_3 $ (and $m_1 = - m_4 $ , $m_2 = - m_5 $ , $m_3 = - m_6 $), but with hierarchical mass-squared differences $ \Delta m_{21}^2 \ll \Delta m_{32}^2  \simeq \Delta m_{31}^2 $, where

\begin{equation}
\Delta m^2_{21} = 1.50 \, \mu^2 \;,\; \Delta m^2_{32} = 425\, \mu^2 \;,\; \Delta m^2_{31} = 427\, \mu^2 
\end{equation}

\ni (and $ \Delta m_{41}^2 = \Delta m_{52}^2 = \Delta m_{63}^2 = 0 $).

The familiar neutrino-oscillation probabilities on the energy shell,

\begin{equation} 
P(\nu_\alpha \rightarrow \nu_\beta) = |\langle \nu_\beta| e^{i PL} |\nu_\alpha \rangle |^2 = \delta _{\beta \alpha} - 4\sum_{j>i} U^*_{\beta j} U_{\beta i} U_{\alpha j} U^*_{\alpha i} \sin^2 x_{ji} 
\end{equation}

\ni with

\vspace{-0.2cm}

\begin{equation} 
x_{ji} = 1.27 \frac{\Delta m^2_{ji} L}{E} \;,\; \Delta m^2_{ji}  = m^2_j - m^2_i \;,\, p_i \simeq E -\frac{m^2_i}{2E} \;,
\end{equation} 

\ni valid when CP violation can be ignored (then $ U_{\alpha i}^* = U_{\alpha i}$), lead  to the formulae

\begin{eqnarray} 
P(\nu_e \rightarrow \nu_e)\, & = & 1 -  \sin^2 x _{21} =  1 - P(\nu_e \rightarrow \nu_\mu)\, - P(\nu_e \rightarrow \nu_\tau)\, \;, \nonumber \\
P(\nu_\mu \rightarrow \nu_\mu) & = & 1 - \frac{1}{4}\sin^2 x _{21} - \frac{1}{2} (\sin^2 x_{31} + \sin^2 x_{32}) = P(\nu_\tau \rightarrow \nu_\tau) = 1 - P(\nu_\mu \rightarrow \nu_\tau)\, \nonumber \\
P(\nu_\mu \rightarrow \nu_e) & = & \frac{1}{2} \sin^2 x _{21} = P(\nu_\tau \rightarrow \nu_e)\,, 
\end{eqnarray} 

\ni when the matrix elements $ U_{\alpha i}$ are calculated from Eqs. (6) and (7) [see  Eq. (A.2)]. Here, the relations $m_1 = - m_4 $ , $m_2 = - m_5 $ , $m_3 = - m_6 $ work in $x_{ji}$ $(i,j =1,2,3,4,5,6)$.

Note that the oscillation formulae (17) hold also in the $3\times 3$ texture of active neutrinos, if the $3\times 3$ mixing matrix $ U = \left(U_{\alpha i}\right)\;\;(\alpha = e\,,\,\mu\,,\,\tau\;,\; i =1,2,3)$  is put equal to the bimaximal mixing matrix $ U^{(3)} = \left( U^{(3)}_{\alpha i} \right)$ given in Eq. (7), while the (conventional) sterile neutrinos are to be decoupled through the familiar seesaw mechanism. In this case, for any mass spectrum $ m_1\,,\,m_2\,,\,m_3 $ the $3\times 3$ mass matrix $ M = \left( M_{\alpha \beta}\right)$ becomes

\begin{equation} 
 M = U^{(3)} \left(\begin{array}{ccc} m_{1} & 0 & 0 \\ 0 & m_{2} & 0 \\ 0 & 0 & m_{3} \end{array}  \right) U^{(3) \dagger} = \left(\begin{array}{ccc} \frac{m_1+ m_2}{2} & \frac{-m_1+ m_2}{2\sqrt{2}} & \frac{m_1- m_2}{2\sqrt{2}} \\ \frac{-m_1+ m_2}{2\sqrt{2}}  & \frac{m_1+ m_2 +2m_3}{4} & \frac{-m_1- m_2 +2m_3}{4} \\ \frac{m_1- m_2}{2\sqrt{2}} & \frac{-m_1- m_2 +2m_3}{4} & \frac{m_1+ m_2 +2m_3}{4} \end{array}  \right) \;. 
\end{equation} 

\ni In particular, for ${\stackrel{0}{m}} = m_1 \simeq m_2 \simeq m_3$ one obtains $M \simeq {\stackrel{0}{m}}\, 1^{(3)}$. The reason, why in both cases the oscillation formulae for active neutrinos $\nu_e \,,\,\nu_\mu \,,\,\nu_\tau $ get the same forms (17), is -- in our texture -- the absence of oscillations for the (conventional) sterile neutrinos $\nu_{e_s} \,,\,\nu_{\mu_s} \,,\,\nu_{\tau_s}$. In fact, one can show from Eqs. (15) that the matrix elements $ U_{\alpha i}$ evaluated with the use of Eqs. (6) and (7) [see Eq. (A.2)] lead to $P(\nu_\alpha \rightarrow \nu_{\beta_s}) = 0$ and $P(\nu_{\alpha_s} \rightarrow \nu_{\beta_s}) = \delta_{\beta_s\,\alpha_s} \;\;(\alpha\,,\,\beta = e\,,\,\mu\,,\,\tau)$ in consequence of $ \Delta m_{41}^2 = \Delta m_{52}^2 = \Delta m_{63}^2 = 0 $, where $m_1 = -m_4\;,\;m_2 = -m_5\;,\;m_3 = -m_6 $.

The oscillation formulae (17) provide the relations

\begin{eqnarray} 
P(\nu_e \rightarrow \nu_e)_{\rm sol}\;\;\;\, & = & 1 - \sin^2 (x _{21})_{\rm sol}\;, \nonumber \\
P(\nu_\mu \rightarrow \nu_\mu)_{\rm atm} \;\; & = & 1 - \frac{1}{4} \sin^2 (x _{21})_{\rm atm} - \frac{1}{2}\left[\sin^2 (x_{31})_{\rm atm} + \sin^2 (x_{32})_{\rm atm}\right] \nonumber \\
& \simeq & 1 - \sin^2 (x_{32})_{\rm atm} \;, \nonumber \\
P(\bar{\nu}_e \rightarrow \bar{\nu}_e)_{\rm Chooz} & = & 1 - \sin^2 (x _{21})_{\rm Chooz} \simeq 1\;, \nonumber \\
P(\nu_\mu \rightarrow \nu_e)_{\rm LSND} & = & \frac{1}{2} \sin^2 (x _{21})_{\rm LSND} \simeq 0\;, 
\end{eqnarray} 

\ni implying {\it bimaximal mixing} for solar $\nu_e$'s and atmospheric $\nu_\mu$'s, {\it negative result} for Chooz reactor $\bar{\nu}_e$'s [6] and {\it no} LSND {\it effect} for  accelerator $\nu_\mu$'s (and $ \bar{\nu}_\mu $'s) [7]. 

From the second formula (19) decribing atmospheric $\nu_\mu$'s we infer due to the SuperKamiokande results [8] that

\begin{equation} 
1 = \sin^2 2\theta_{\rm atm} \sim 1 \;\;,\;\; \Delta m^2_{32} = \Delta m^2_{\rm atm} \sim 3 \times 10^{-3}\;\,{\rm eV} \;,
\end{equation} 

\ni what gives 

\begin{equation} 
\mu^2 \sim 7.1 \times 10^{-6} \; {\rm eV}^2\;\, {\rm or}\,\;\mu \sim 2.7 \times 10^{-3} \; {\rm eV}  \;,
\end{equation} 

\ni when Eq. (14) is used. 

Then, the first formula (19) referring to solar $\nu_e$'s {\it predicts}

\begin{equation} 
\sin^2 2\theta_{\rm sol} = 1 \;\,,\,\; \Delta m^2_{\rm sol} = \Delta m^2_{21} \sim 1.1 \times 10^{-5}\;\,{\rm eV}^2 \;,
\end{equation} 

\ni where Eqs. (14) and (21) are applied. Such a prediction for solar $\nu_e$'s is not inconsistent with the Large Mixing Angle (LMA) solar solution [9].

\vspace{0.2cm}

{\bf 4. Conclusions. } We presented in this note an effective texture for six Majorana neutrinos, three active and three (conventional) sterile, based on the $6\times 6$ mass matrix defined in Eqs. (3) and (5), and leading to the mixing matrix given in Eqs. (6) and (7), as well as to the mass spectrum (8) or (9). We conjectured that the Dirac $3\times 3$ component of such a neutrino  mass matrix, when the bimaximal mixing specific for neutrinos is transformed out unitarily, gets a {\it fermion universal form} (11) similar to the $3\times 3 $ mass matrix for charged leptons and $3\times 3$ mass matrices for up and down quarks, constructed previously with a considerable success [2,3].

This texture {\it predicts} reasonably oscillations of solar $\nu_e$'s in a form not inconsistent with  LMA solar solution, if the SuperKamiokande value of the mass--squared scale for atmospheric $\nu_\mu $'s is taken as an input. In both cases, neutrino oscillations are maximal. The proposed texture {\it predicts} also the negative result of Chooz experiment for reactor $\bar{\nu}_e$'s and the absence of LSND effect for accelerator $\nu_\mu$'s (and $\bar{\nu}_\mu $'s). The new miniBooNE experiment may confirm or revise the original LSND results.

As far as the neutrino mass spectrum is concerned, our model of neutrino texture is of 2+2+2 type (with $m_1 = | m_4 | \stackrel{<}{\sim} m_2 = | m_5 | \stackrel{<}{\sim} m_3 = | m_6 |$), in contrast to the models of 3+1 or 2+2 types [10] discussed in the case when, beside three active neutrinos $\nu_e\,,\,\nu_\mu\,,\,\nu_\tau $, there is one {\it extra} sterile neutrino $\nu_s $. In those models, three Majorana {\it conventional} sterile neutrinos $\nu_{e_s}\,,\, \nu_{\mu_s} \,,\, \nu_{\tau_s}$ are decoupled through the familiar seesaw mechanism, as being practically identical with three very heavy mass neutrinos $\nu_4\,,\,\nu_5\,,\,\nu_6 $ (of the GUT mass scale). In our model, on the contrary, the mass neutrinos $\nu_4\,,\,\nu_5\,,\,\nu_6 $ are constructed to be degenerate in mass magnitude with the light mass neutrinos $\nu_1\,,\,\nu_2 \,,\,\nu_3 $, respectively. In consequence of this degeneracy, the conventional sterile neutrinos   $\nu_{e_s}\,,\, \nu_{\mu_s} \,,\,\nu_{\tau_s}$ do not oscillate in our neutrino texture.

\vfill\eject
~~~

{\centerline{\bf Appendix}}

\vspace{0.3cm}

In our neutrino texture, the explicit forms of the neutrino mass matrix and its diagonalizing matrix are

$$
M = {\stackrel{0}{m}}\left( \begin{array}{cccccc} 1 & 0 & 0 & \frac{\tan 2\theta_{14} }{\sqrt2}& \frac{\tan 2\theta_{25}}{\sqrt{2}} & 0  \\ \\ 0 & 1 & 0 & -\frac{\tan 2\theta_{14}}{2} & \frac{\tan 2\theta_{25}}{2} & \frac{\tan 2\theta_{36} }{\sqrt2} \\ \\ 0 & 0 & 1 & \frac{\tan 2\theta_{14}}{2} & -\frac{\tan 2\theta_{25}}{2} & \frac{\tan 2\theta_{36}}{\sqrt2} \\ \\ \frac{\tan 2\theta_{14}}{\sqrt2} & -\frac{\tan 2\theta_{14}}{2} & \frac{\tan 2\theta_{14}}{2} & -1 & 0 & 0 \\ \\ \frac{\tan 2\theta_{25}}{\sqrt2} & \frac{\tan 2\theta_{25}}{2} & -\frac{\tan 2\theta_{25}}{2} & 0 & -1 & 0 \\ \\ 0 & \frac{\tan 2\theta_{36}}{\sqrt2} & \frac{\tan 2\theta_{36}}{\sqrt2} & 0 & 0 & -1 \end{array} \right)
\eqno({\rm A}.1)
$$

\vspace{0.2cm}

\ni [from Eqs(3) and (5)] and

$$
U =  \left( \begin{array}{cccccc} \frac{c_{14}}{\sqrt{2}} & \frac{c_{25}}{\sqrt{2}} & 0 & -\frac{s_{14}}{\sqrt{2}} & -\frac{s_{25}}{\sqrt{2}} & 0  \\ \\ -\frac{c_{14}}{2} & \frac{c_{25}}{2} & \frac{c_{36}}{\sqrt{2}} & \frac{s_{14}}{2} & -\frac{s_{25}}{2} & -\frac{s_{36}}{\sqrt{2}} \\ \\ \frac{c_{14}}{2} & -\frac{c_{25}}{2} & \frac{c_{36}}{\sqrt{2}} & -\frac{s_{14}}{2} & \frac{s_{25}}{2} & -\frac{s_{36}}{\sqrt{2}}\\ \\ s_{14} & 0 & 0 & c_{14} & 0 & 0 \\ \\ 0 & s_{25} & 0 & 0 & c_{25} & 0 \\ \\ 0 & 0 & s_{36} & 0 & 0 & c_{36} \end{array} \right) 
\eqno({\rm A}.2)
$$

\vspace{0.2cm}

\ni [from Eqs. (6) and (7)], respectively, where the neutrino mass
spectrum is given as $ m_{i,j} = \pm {\stackrel{0}{m}}\sqrt{1 + \tan^2
2\theta_{ij}}$ or $(c^2_{ij} - s^2_{ij}) m_{i,j} = \pm
{\stackrel{0}{m}}$ with $c_{ij} = \cos \theta_{ij}$ and $s_{ij} = \sin
\theta_{ij}\;\,(ij = 14,25,36)$. Of course, $ M_{\alpha \beta} =
\sum_i U_{\alpha i} m_i U^*_{\beta i}$ and $\sum_{\alpha \beta}
U^*_{\alpha i} M_{\alpha \beta} U_{\beta j} = m_i \delta_{i
j}\;\, $ $ (\alpha,\beta =  e\,,\, \mu\,,\, \tau\,,\,e_s\,,\, \mu_s\,,\,
\tau_s\;,\; i,j = 1,2,3,4,5,6) $. Here, $\tan 2\theta_{14} \rightarrow
0$ , $\tan 2\theta_{25} = 1.23 \mu/{\stackrel{0}{m}}$ and $\tan
2\theta_{36} = 20.7\mu/{\stackrel{0}{m}}$ [from Eqs. (12)] with $\mu
\sim 2.7 \times 10^{-3}$ eV [from Eq. (21)]. 

The Dirac component $M^{(D)}$ of the mass matrix $M$, breaking the electroweak symmetry, may be of the Higgs origin. On the contrary, the lefthanded and righthanded components $M^{(L)}$ and $M^{(R)}$ of $M$, the former breaking the electroweak symmetry, may be given explicitly. The bimaximal mixing matrix 

$$
U^{(3)} = \left( \begin{array}{rrr}  \frac{1}{\sqrt2} & \frac{1}{\sqrt2} & 0 \\ -\frac{1}{2} & \frac{1}{2} & \frac{1}{\sqrt2} \\ \frac{1}{2} & -\frac{1}{2} & \frac{1}{\sqrt2} \end{array} \right) \;, 
\eqno({\rm A}.3)
$$

\ni operating within the Dirac component $M^{(D)} = {\stackrel{0}{m}}\,U^{(3)} \,{\rm diag}(\tan 2\theta_{14}\,,\, \tan 2\theta_{25}\,,\, \tan 2\theta_{36})$ of $M$ as a factor and so, breaking the electroweak symmetry by itself, may be introduced also explicitly. Note that $ U^{(3)}$ is the generic $ 3\times 3$ mixing matrix with $\theta_{12} = \pi/4\;,\;\theta_{23} = \pi/4 $ and $\theta_{13} = 0$. Deforming unitarily $\,{\stackrel{0}{m}}\,{\rm diag}(\tan 2\theta_{14}\,,\, \tan 2\theta_{25}\,,\, \tan 2\theta_{36})$, the matrix $ U^{(3)}$ introduces the significant difference between the hierarchical structures of neutrinos and charged leptons. 

From Eqs. (2) and (A.2) one can see that the flavor neutrinos $\nu_\alpha $ are built up as the simplest orthogonal combinations of the superpositions $\nu_{ij} \equiv c_{ij}\nu_i - s_{ij}\nu_j \simeq \nu_i $ and $\nu'_{ij} \equiv s_{ij}\nu_i + c_{ij}\nu_j \simeq \nu_j\,\; (ij = 14,25,36)$ of the mass neutrinos $\nu_i$:

\begin{eqnarray*}
\nu_e & = &  \frac{1}{\sqrt 2}(\nu_{14} + \nu_{25}) \simeq \frac{1}{\sqrt 2}(\nu_1 + \nu_2) \;, \\
\nu_\mu & = & \frac{1}{\sqrt 2} \left[-\frac{1}{\sqrt 2}(\nu_{14} - \nu_{25}) + \nu_{36}\right] \simeq \frac{1}{\sqrt 2}\left[-\frac{1}{\sqrt 2}(\nu_1 - \nu_2) + \nu_3\right]  \;, \\
\nu_\tau & = & \frac{1}{\sqrt 2} \left[\frac{1}{\sqrt 2}(\nu_{14} - \nu_{25}) + \nu_{36}\right] \simeq \frac{1}{\sqrt 2} \left[\frac{1}{\sqrt 2}(\nu_1 - \nu_2) + \nu_3\right]  \;, \\
\nu_{e_s}\! & = & \nu'_{14} \simeq \nu_4 \;, \\
\nu_{\mu_s}\!  & = & \nu'_{25} \simeq \nu_5 \;, \\
\nu_{\tau_s}\!  & = & \nu'_{36} \simeq \nu_6 \;,
\end{eqnarray*}

\vspace{-1.55cm}

\begin{flushright}
({\rm A}.4)
\end{flushright}

\vspace{0.2cm}

\ni where the second step is taken for $s_{ij} \ll c_{ij}$ corresponding to $\mu \ll {\stackrel{0}{m}} $.

In the mathematical limit of $ s_{ij} \rightarrow 0\,\; (ij = 14,25,36) $ corresponding to $ \mu/ {\stackrel{0}{m}} \rightarrow 0 $ one obtains $M \rightarrow {\stackrel{0}{m}}\, {\rm diag}(1^{(3)} \,,\,-1^{(3)}) $ and $U \rightarrow {\stackrel{1}{U}} = {\rm diag}(U^{(3)} \,,\, 1^{(3)})$, where $m_{i,j} \rightarrow \pm {\stackrel{0}{m}} $. Then, together with $ M^{(D)}$, the neutrino hierarchical structure vanishes.

\vfill\eject

~~~~
\vspace{0.5cm}

{\centerline{\bf References}}

\vspace{0.45cm}

{\everypar={\hangindent=0.6truecm}
\parindent=0pt\frenchspacing

{\everypar={\hangindent=0.6truecm}
\parindent=0pt\frenchspacing

~[1]~For a theoretical summary {\it cf.} J.~Ellis, {\it Nucl. Phys. Proc. Suppl.} {\bf 91}, 503 (2001); and references therein.

~[2]~W.~Kr\'{o}likowski, in {\it Spinors, Twistors, Clifford Algebras and Quantum Deformations (Proc. 2nd Max Born Symposium 1992)}, eds. Z.~Oziewicz {\it et al.}, Kluwer Acad. Press, 1993; {\it Acta Phys. Pol.} {\bf B 27}, 2121 (1996); and references therein.

\vspace{0.2cm}

~[3]~W. Kr\'{o}likowski, {\it Acta Phys. Pol.} {\bf B 30}, 2631 (1999); hep--ph/0108157; and references therein.

\vspace{0.2cm}

~[4]~W. Kr\'{o}likowski, hep--ph/0109212; hep--ph/0201004; in both papers the neutrino spectrum is of 3+3 type, in the second with $m_4 = m_5 = m_6 = 0$.

\vspace{0.2cm}

~[5]~The Particle Data Group, {\it Eur. Phys. J.} {\bf C 15}, 1 (2000).

\vspace{0.2cm}

~[6]~M. Appolonio {\it et al.}, {\it Phys. Lett.} {\bf B 420}, 397 (1998); {\bf B 466}, 415 (1999).
\vspace{0.2cm}

~[7]~G. Mills, {\it Nucl. Phys. Proc. Suppl.} {\bf 91}, 198 (2001);  R.L.~Imlay, Talk at {\it ICHEP 2000} at Osaka; and references therein.

\vspace{0.2cm}

~[8]~T. Kajita and Y. Totsuka, {\it Rev. Mod. Phys.} {\bf 73}, 85 (2001); T. Toshito, hep--ex/0105023.

\vspace{0.2cm}

~[9]~For a recent analysis {\it cf.} M.V.~Garzelli and C.~Giunti, hep--ph/010819; hep--ph/0111254; {\it cf.} also V. Barger, D. Marfatia and K. Whisnant, hep--ph/0106207.

\vspace{0.2cm}

[10] {\it Cf. e.g.} V. Barger, B. Kayser, J. Learned, T. Weiler and K. Whisnant, {\it  Phys. Lett.} {\bf B 489}, 345 (2000); M.C. Gonzalez--Garcia, M. Maltoni and C. Pe\~{n}a--Garay, hep--ph/0108073; and references therein; {\it cf}. also W. Kr\'{o}likowski, hep--ph/0106350; O.Yasuda, hep--ph/0109067.

\vfill\eject

\end{document}